\documentclass[%
 reprint,
 amsmath,amssymb,
 aps,
]{revtex4-2}
  
\usepackage{graphicx}
\usepackage{dcolumn}
\usepackage{bm}

\usepackage[caption=false]{subfig}

\usepackage{url}
\usepackage{braket}
\usepackage{hyperref}
\hypersetup{
  colorlinks   = true,    
  urlcolor     = blue,    
  linkcolor    = blue,    
  citecolor    = blue,    
}

\usepackage{amsmath} 

\begin{document}

\preprint{APS/123-QED}

\title{The impact of nitrogen doping on the linear and nonlinear terahertz response of graphene } 

\author{Roozbeh Anvari}
\author{Eugene Zaremba}
\author{Marc M. Dignam}
\email{dignam@queensu.ca}
\affiliation{Department of Physics, Engineering Physics \& Astronomy,\\ Queen’s University, Kingston, Ontario K7L 3N6, Canada.}

\date{\today} 
\begin{abstract}
 It is well known that impurities play a central role in the linear and nonlinear response of graphene at optical and terahertz frequencies. In this work, we calculate the bands and intraband dipole connection elements for nitrogen-doped monolayer graphene using a density functional tight binding approach. Employing these results, we calculate the linear and nonlinear response of the doped graphene to terahertz pulses using a density-matrix approach in the length gauge. We present the results for the linear and nonlinear mobility as well as third harmonic generation in graphene for adsorbed and substitutional nitrogen doping for a variety of doping densities.  
 We show that the conduction bands are more parabolic in graphene structures with substitutional nitrogen doping than for those with adsorbed nitrogen. As a result, substitutional doping has a greater impact on the terahertz mobility and nonlinear response of graphene than adsorbed nitrogen does.
\end{abstract}

\maketitle


\section{Introduction}

Graphene, a monolayer of $sp^2$-bonded carbon atoms, has attracted much interest due to its extraordinary characteristics. The high mobility of graphene makes it a perfect platform for realizing high speed devices, such as graphene-based terahertz-frequency switches. \cite{lee2012switching_switch, li2018graphene_switch}
Linear dispersion near the Dirac points and the tunability of the chemical potential through an applied gate voltage \cite{shi2014controlling, novoselov2004electric} are few of the remarkable features that make graphene an attractive material system for nonlinear optics and harmonic generation \cite{mikhailov2007non_cvd,hafez2020terahertz_THG}. Both chemical vapour deposition \cite{horng2011drude} and epitaxial graphene growth \cite{choi2009broadband} techniques have matured since graphene's emergence in 2004 \cite{novoselov2004electric}. Although there are many theoretical papers on the linear and nonlinear optical and terahertz response of pristine graphene, \cite{Marc_al2015nonperturbative, Marc_Ibr_2015optimizing, Marc_Luke_2019effect, malic2017carrier} 
to our knowledge, a comprehensive theoretical examination of carrier dynamics in \textit{non-pristine} graphene \cite{johnston_docherty2012extreme, exp_knight2017situ, exp_N_gr_lu2013nitrogen} has not yet been carried out. In this work we examine the impact of nitrogen doping on the linear and nonlinear response of graphene to terahertz (THz) radiation.\\

Various theoretical and experimental studies have been performed to gain a better understanding of carrier relaxation dynamics in graphene \cite{wang2010ultrafast, dawlaty2008measurement, knorr_kadi2015impact, winnerl2011carrier}. However, the effect of defects on carrier dynamics, in particular in the terahertz regime, is not yet well understood. It is known  that short range scatterers (e.g. lattice defects) and long range scatterers (e.g. ionized impurities or ripples) cause intervalley, and intravalley scattering \cite{exp_N_gr_lu2013nitrogen, gr_scatter_castro2009electronic, gr_scatter_peres2010colloquium}, which emphasizes the significant role of disorder in graphene. In addition to intrinsic defects, oxygen, nitrogen, and water are among the important impurities that are often intentionally or unintentionally introduced to graphene \cite{johnston_docherty2012extreme, exp_knight2017situ, hafez2017Dopingeffects}. Nitrogen-doped graphene has found applications in biosensors \cite{wang2010nitrogen_N_gr_fab}, field effect transistors \cite{zhang2011synthesis_N_gr_fab}, batteries \cite{reddy2010synthesis_N_gr_fab}, and supercapacitors \cite{jeong2011nitrogen_N_gr_fab}. Nitrogen can also appear as a contaminant in processes such as chemical vapor deposition of methane in the presence of ammonia \cite{wei2009synthesis_N_gr_fab}, thermal annealing of graphene oxide in ammonia \cite{li2009simultaneous_N_gr_fab} and nitrogen plasma treatment of graphene \cite{wang2010nitrogen_N_gr_fab}. In addition, it can arise simply due to exposure to air  \cite{johnston_docherty2012extreme}.  

 Theoretical and experimental studies have shown that doping graphene can significantly alter its physical and chemical properties \cite{exp_knight2017situ, johnston_docherty2012extreme}. It is also known that nitrogen doping can cause a band gap to open and thereby introduce semiconducting properties to graphene \cite{nath2014ab, usachov2011nitrogen_fab, lherbier2013electronic, xiang2012ordered, rani2013designing}. More generally, nitrogen doping alters the electronic structure, carrier density, and linear and nonlinear conductivity of graphene. Pump-probe experiments of Docherty \textit{et al.} \cite{johnston_docherty2012extreme} showed that exposure of graphene to oxygen and nitrogen gases significantly changes the differential terahertz response of graphene. They also found that exposure of graphene to air causes a differential response that is similar to that of graphene being exposed to a mixture of oxygen and nitrogen gases, and that the adsorption of either of these gases is molecular and reversible when the system is re-evacuated. However, the microscopic origin of these experimentally-observed effects is not well understood. \\  
 
Various models of the effects of impurities on carrier dynamics in graphene driven by terahertz radiation can be found in the literature. Some include impurities phenomenologically via scattering times but retain the linear energy dispersion curves of pristine graphene. \cite{knorr_kadi2015impact, TB_dopedGraphene_leconte2010damaging, Marc_Luke_2019effect, Sipe_cheng2013optical_LDA}. Such approaches neglect the effect of defects on the band structure, such as the opening of a band gap. On the other hand \textit{ab initio} studies have mainly focused on the band structure \cite{tran2017coverage_DFT}, material characteristics at optical frequencies \cite{rani2014dft, johari2011modulating_DFT, muhammad2017first_DFT_tr}, or carrier dynamics for low-intensity incident fields \cite{rigo2009electronic_DFT_tr}. Previous work in our group focused on the linear and nonlinear response of monolayer and bilayer graphene to terahertz radiation, and the interplay between carrier dynamics, phonon scattering and neutral impurity scattering using a microscopic scattering model, but neglected impurity-induced changes in the electronic structure \cite{hafez2017Dopingeffects, Marc_Luke_2019effect, Marc_Ibr_2015optimizing, Marc_al2015nonperturbative}. In the present work we combine a density functional tight binding method along with a density matrix treatment of carrier dynamics and phenomenological scattering to study the effects of nitrogen doping on the linear and nonlinear THz response of graphene.\\

The paper is organized as follows. In section \ref{Theoretical model} we present our density matrix formulation of carrier dynamics for a single band. In section \ref{atomic structure}, we present a detailed analysis of the effects of nitrogen doping on the atomic coordinates and energetics of the nitrogen-doped structures we have studied. Corresponding band structures, velocities, connection elements, and Berry curvatures are discussed in section \ref{res_energy_bands}. In section \ref{carrier dynamics}, we use our model to simulate carrier dynamics, and present results for intraband currents, the transmitted field, linear and nonlinear mobility, and harmonic generation in different nitrogen-doped graphene systems as a function of the incident field, chemical potential and defect density. Finally, we conclude in section \ref{conclusion}.\\

\section{Theoretical model} \label{Theoretical model}

We use a density functional tight-binding (DFTB) code to calculate the relaxed atomic coordinates, energy dispersion curves, and the corresponding Bloch states for a number of different periodic, nitrogen-doped graphene structures. We employ DFTB because it is a very efficient method by which one can obtain analytic expressions for the Bloch states. This is critical, as it gives a computationally-efficient way to obtain the band structure and the interband and intraband connection elements. All the calculated parameters are then used in the dynamic equations to calculate the time evolution of charged carriers in the conduction band in response to pulsed THz fields. From these calculations, we extract the linear and nonlinear mobility as well as the generated third harmonic field. \\ 

To model the effects of N-doping on the response of graphene, we construct $(5 \times 5)$ graphene supercells with one, two, or three N defects per unit cell, which allows us to model different doping densities. We then use self-consistent-charge DFTB (SCC-DFTB) to calculate the Bloch states and energy bands for each of these supercells.  We start by discussing the Bloch states and their use in calculating the parameters needed in our dynamic calculations.  \\

The tight binding expression for the Bloch state, $\psi_{n\textbf{k}}(\textbf{r}) \equiv\langle \textbf{r}|n,\textbf{k}\rangle$ for band $n$ and wave vector $\textbf{k}$ is given by
\begin{equation}
\label{eq_1}
\psi_{n\textbf{k}}(\textbf{r}) = \sum_{\textbf{R}} \sum_{j=1}^{n_B} C_j^n(\textbf{k}) \phi_j(\textbf{r} -\textbf{r}_j -\textbf{R}) e^{i\textbf{k} \cdot \textbf{R}},  
\end{equation}
where $\textbf{R}$ are the two-dimensional Bravais lattice vectors and the sum over $j$ runs over all $n_B$ outer-shell $s$ and $p$ orbitals $\phi_{j}(\textbf{r})$ of all carbon and nitrogen atoms at positions $\textbf{r}_{j}$ in the super cell. The  $C^n_{j}(\textbf{k})$ are the sublattice expansion coefficients. 

We use the DFTB+ code \cite{aradi2007dftb+, koskinen2009density, dftb_hourahine2020} to self-consistently solve the general eigenproblem, 
\begin{equation}
\label{eq_2}
\sum_{j^{'}} H_{jj^{'}}(\textbf{k})C^{n}_{j^{'}}(\textbf{k}) = E_{n}(\textbf{k}) \sum_{j^{'}} S_{jj^{'}}(\textbf{k}) C^{n}_{j^{'}}(\textbf{k}),  
\end{equation}
for the fully relaxed supercell to obtain the expansion coefficients $C^{n}_j (\textbf{k})$ and band energies $E_n(\textbf{k})$. Here
\begin{equation}
S_{jj^{'}}(\textbf{k})  \equiv \sum_{\textbf{R}}e^{i\textbf{k} \cdot \textbf{R}}\int_{V} d^3\textbf{r}\phi_{j}^{*}(\textbf{r}-\textbf{r}_{j})\phi_{j^{'}}(\textbf{r}-\textbf{r}_{j^{'}}-\textbf{R}),  
\end{equation}
are overlap integrals between Bloch states formed from different orbitals, and $H_{jj^{'}}(\textbf{k})$ are the matrix elements of the electron Hamiltonian between Bloch states formed from different atomic orbitals. The latter can be broken up into two terms as
\begin{equation}
H_{jj^{'}}(\textbf{k}) = H^0_{jj^{'}}(\textbf{k}) + h^1_{jj^{'}}(\textbf{k}) S_{jj^{'}}(\textbf{k}) 
\end{equation}
where $H^0_{jj^{'}}(\textbf{k})$ are the matrix elements of the non-self-consistent Hamiltonian $H^0$, which depends on the inter-atomic distances and reference densities of neutral atoms in their geometry inside the lattice \cite{gaus2011dftb3}:

\begin{equation}
H^0_{jj^{'}}(\textbf{k}) \equiv \sum_{\textbf{R}}e^{i\textbf{k} \cdot \textbf{R}}\int_{V} d^3\textbf{r}\phi_{j}^{*}(\textbf{r}-\textbf{r}_{j})H^0\phi_{j^{'}}(\textbf{r}-\textbf{r}_{j^{'}}-\textbf{R}),
\end{equation}
 and the matrix elements  $h^1_{jj^{'}}(\textbf{k})$ add a correction to $H_{jj^{'}}(\textbf{k})$ arising from the averaged self-consistently-calculated electrostatic potential around orbitals $j$ and $j^{'}$ \cite{aradi2007dftb+, koskinen2009density}. The integrals found in the expressions for $ H^0_{jj^{'}}(\textbf{k})$ and  $S_{jj^{'}}(\textbf{k})$ are obtained from DFTB parameterization files \cite{dftb_mio_elstner1998self}, and Slater-Koster orientation rules are applied for the actual orientation of the $j$-$j^{'}$ pair of orbitals \cite{slater_Koster_orientation_1954simplified}.  \\
 
For the Bloch states, we apply the normalization condition
\begin{equation}
\label{eq_6}
\int_{\Omega_0} d^3\textbf{r}\, u^*_{n\textbf{k}}(\textbf{r})u_{m\textbf{k}}(\textbf{r}) = \frac{\Omega}{(2\pi)^2} \delta_{nm}, \\                    
\end{equation}
where 
\begin{equation} 
\label{eq_7}
u_{m\textbf{k}}(\textbf{r}) = e^{-i\textbf{k} \cdot \textbf{r}} \psi_{m\textbf{k}}(\textbf{r}),  
\end{equation}
is the periodic part of the Bloch state, $\Omega$ is the 2D supercell area, and $\Omega_0$ is the volume of a supercell (\textit{i.e.} the 2D supercell extended above and below the plane of the graphene). The above equation results in the following normalization condition for the expansion coefficients (see Appendix \ref{Appendix A}):
\begin{equation}
\sum_{jj^{'}} C^{n*}_j(\textbf{k}) C^{m}_{j^{'}}(\textbf{k}) S_{jj^{'}}(\textbf{k}) = \frac{\Omega}{(2\pi)^2}\delta_{nm}. 
\end{equation}

The field-carrier interaction is treated in the length gauge so that divergences at low frequencies are avoided \cite{al2014high_gaug, aversa_sipe_1995}. In this gauge, the full Hamiltonian is given by $H^{'}=H-e\textbf{r}\cdot\textbf{E}_t(t)$, where $e= -|e|$ is the charge on an electron, $\textbf{r}$ is the position operator of the electron, and $\textbf{E}_t(t)$ is the transmitted THz electric field in the plane of the graphene. 

The carrier dynamics can be calculated by solving the equations of motion for the reduced density matrix in the Bloch-state basis, in which the matrix elements of the Hamiltonian are 
\begin{equation}
\langle n\textbf{k} \left| H \right|  m\textbf{k}' \rangle = E_n(\textbf{k})\delta_{nm}\delta(\textbf{k}   -\textbf{k}' ) - e \langle n\textbf{k} \left| \textbf{r} \right|  m\textbf{k}' \rangle \cdot \textbf{E}_{t}(t),
\end{equation}
where the matrix elements of $\textbf{r}$ between Bloch states are given by \cite{aversa_sipe_1995} 
\begin{equation}
\langle n\textbf{k} \left| \textbf{r} \right|  m\textbf{k}' \rangle = \delta(\textbf{k}   -\textbf{k}' )  \boldsymbol{\xi}_{nm}(\textbf{k}) + i\delta_{nm} \boldsymbol{\nabla}_{\textbf{k}} \delta(\textbf{k}   -\textbf{k}' ).
\end{equation}
Here, the connection elements $\boldsymbol{\xi}_{nm}(\textbf{k})$ are defined as 
\begin{equation}
\label{eq_11}
\boldsymbol{\xi}_{nm}(\textbf{k}) =\frac{i(2\pi)^2}{\Omega} \int_{\Omega_0} d^3\mathbf{r} u^*_{n\textbf{k}}(\mathbf{r}) \boldsymbol{\nabla}_{\textbf{k}} u_{m\textbf{k}}(\mathbf{r}). 
\end{equation}

In our previous work, we employed a nearest-neighbour tight binding (NNTB) model in which the overlap of atomic wave functions was neglected. \cite{Marc_Luke_2019effect, Marc_Ibr_2015optimizing, Marc_al2015nonperturbative} However, in this work we go beyond nearest neighbours. As is shown in Appendix \ref{Appendix B}, accounting for atomic overlaps results in the following expression for the connection elements: 
\begin{equation}
\label{eq_12}
\boldsymbol{\xi}_{nm}(\textbf{k}) = \frac{i(2\pi)^2}{\Omega} \sum_{jj^{'}}  C_{j}^{n*}(\textbf{k})  \left[\boldsymbol{\nabla}_\mathbf{k}  C_{j^{'}}^{m}(\textbf{k})\right]
S_{jj^{'}}(\textbf{k}).
\end{equation}

The dynamic equations for the matrix elements $\rho_{nm}(\textbf{k}) =\langle n,\textbf{k}|\hat{\rho}|m,\textbf{k}\rangle$ of the density operator $\hat{\rho}(t)$ for carriers in the {\it n}-th band are \cite{aversa_sipe_1995}
\begin{equation}
\label{eq_13}
\begin{split}
\frac{d\rho_{nn}(\textbf{k})}{dt} & = 
-\frac{2e\textbf{E}_t(t)}{\hbar}\cdot \sum_{m,m\neq n} \rm Im[\boldsymbol{\xi}_{nm}(\textbf{k})\rho_{mn}(\textbf{k})] \\
&-e\frac{\textbf{E}_t(t)}{\hbar}\cdot \boldsymbol{\nabla}_{\textbf{k}}\rho_{nn}(\textbf{k}) -\frac{[\rho_{nn}(\textbf{k}) -f(E_n(\textbf{k}))]}{\tau_n} \,. 
\end{split} 
\end{equation}
Here, $f(E)$ is the Fermi-Dirac distribution and $\tau_n$ is a phenomenological scattering time that arises from various scattering mechanisms such as neutral impurities, acoustic and optical phonons, and substrate charged impurities \cite{Marc_Luke_2019effect}. 
At room temperature, scattering times in graphene are on the order of tens of femtosecconds \cite{al2014high_gaug, exp_thz_paul2013high}, which necessitates the inclusion of scattering processes in carrier dynamics equations. Although N-doping will in principle increase the scattering rate, we assume that the scattering is dominated by phonons and other impurities so that the scattering times $\tau_n$ are independent of the doping density. \\

There is a similar equation of motion for the off-diagonal elements of the density operator. However, in all of our simulations, we take the chemical potential ($\mu_c$) to be sufficiently high (100 meV and above) such that the terahertz pulses do not excite interband transitions, and so the only contribution to the current density comes from intraband motion. Thus, Eq. (\ref{eq_13}) need only be solved for the conduction band, with $\rho_{mc}(\textbf{k})=0$ for $m\neq c$. A finite difference approximation to the gradients is applied and a fourth-order Runge-Kutta method is used to solve the above equations on a hexagonal grid uniformly sampled in k-space about the $K$ and $K'$ points \cite{Marc_al2015nonperturbative}.

Following the formalism of Aversa and Sipe \cite{aversa_sipe_1995}, the current density is given by 
\begin{equation}
\label{eq_14}
\textbf{J}(t) = \frac{e}{mA}Tr\{\textbf{p}\hat{\rho}(t)\}=\frac{e}{i\hbar A} Tr\{ [\textbf{r}, H]\hat{\rho}(t) \},
\end{equation}
where the trace is over single electron states, $A$ is the area of the graphene sheet, and \textbf{p} is the electron momentum operator. It can be shown that using our expression for the matrix elements of the Hamiltonian and position operator, the current density is given by \cite{Marc_al2015nonperturbative} 
\begin{equation} 
\label{eq_15}
\textbf{J}(t)  = \frac{2e}{A\hbar} \sum_\textbf{k} \rho_{cc}(\textbf{k}) 
\left\{\boldsymbol{\nabla}_\mathbf{k} E_c(\textbf{k}) 
-e \mathbf{E}_t(t) \times \boldsymbol{\Omega}_B (\textbf{k})  \right\}  \\,
\end{equation}
where $\boldsymbol{\Omega}_B (\textbf{k}) \equiv \boldsymbol{\nabla}_\mathbf{k} \times\boldsymbol{\xi}_{cc}(\textbf{k}) $ is the Berry curvature for the conduction band, which points perpendicular to the plane of the graphene. In our numerical simulations the sums over \textbf{k} are restricted to the areas close to the two Dirac points, $K$ and $K^{'}$ where $E_c(\textbf{k})-E_{min} \leq 700 $ meV. As will be discussed in the results section, due to the asymmetry of the bands near $K$ and $K^{'}$ in defected structures, the integration is performed over both the $K$ and $K^{'}$ valleys (unlike our previous work on pristine graphene). The factor of 2 is included to account for spin degeneracy. We note that in pristine graphene, the Berry curvature is zero at all points in $k$-space, except right at $K$ and $K^{'}$ where it is not defined. Thus, the ``anomalous" current arising from the second term in Eq. (\ref{eq_15}) is absent in pristine graphene. However, as we shall show, in N-doped graphene there can be significant regions in $k$-space near $K$ and $K^{'}$ over which the Berry curvature is non-negligible. As a result, the second term in Eq. (\ref{eq_15}) can result in significant anomalous currents in the $K$ and $K^{'}$ valleys, which are equal in magnitude, but opposite in sign. These are the so-called ``valley-Hall" currents \cite{valHall_friedlan2021valley, valHall_cresti2016charge}. \\

We assume that the graphene sheet is at the interface between air and a substrate with refractive index $n$. Then it can be shown that satisfying the boundary conditions for the fields gives the transmitted field at the position of the graphene:
\begin{equation}
\label{eq_16}
\textbf{E}_t(t) = \frac{2 \textbf{E}_i(t)- Z_\circ \textbf{J} [ \textbf{E}_t(t) ] }{1+n},
\end{equation}
where $\textbf{E}_i(t)$ is the incident terahertz field, $Z_\circ$ is the impedance of free space, and $\textbf{J} [ \textbf{E}_t(t) ]$ is the graphene current density calculated using the transmitted field as the driving field. This equation is used to iteratively calculate the transmitted terahertz field at each time step \cite{Marc_al2015nonperturbative}. 

\section{Results} \label{results}

In this section, we present the results of SCC-DFTB calculations of the atomic structures, band structures and the THz response of a number of different N-doped graphene systems. The structures considered are those we believe to be the most likely to occur. The atomic coordinates and energetics of these structures are summarized in section \ref{atomic structure}. The calculated band structures, electron velocities, connection elements, and Berry curvatures are discussed in section \ref{res_energy_bands}. Finally, the linear and nonlinear terahertz response of these structures is discussed in section \ref{carrier dynamics}.  

\subsection{Atomic structure} \label{atomic structure}
The DFTB+ package is used to determine the relaxed atomic coordinates with a k-mesh of 20 $\times$ 20 $\times$ 1. For geometry optimization, a conjugate-gradient algorithm was used, imposing a maximum force difference of  $1.0 \times 10^{-4} \: \rm eV/\rm \AA$ and a maximum self-consistent charge tolerance of $1.0 \times 10^{-5} \: \rm |e|$ \cite{aradi2007dftb+}. Nitrogen and carbon interactions are described using the parameterization (Slater-Koster tables) developed by Elstner \textit{et al.} \cite{dftb_mio_elstner1998self}. The fully relaxed lattice parameters of the undoped $(5\times5)$ supercell are $a/5$=2.47136 \rm \AA, $b/5$=2.47134 \rm \AA, $\gamma$=$120^\circ$, and the graphene sheet is separated from its periodic image by a 15 $\rm \AA$ vacuum gap. \\

\begin{figure}
\centering
\subfloat[\label{fig_1_a}]{%
  \includegraphics[width=0.5\columnwidth]{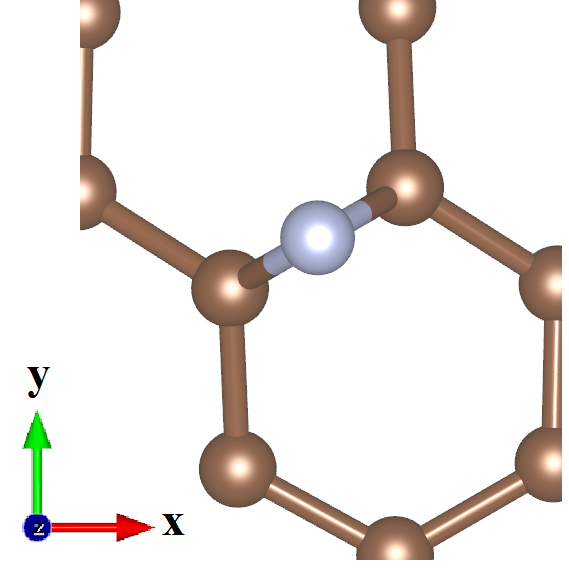} %
  }    
\subfloat[\label{fig_1_b}]{%
  \includegraphics[width=0.5\columnwidth]{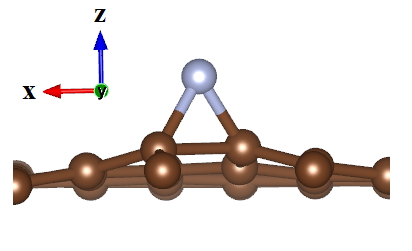} 
  } \hfill
\subfloat[\label{fig_1_c}]{%
  \includegraphics[width=0.3\columnwidth]{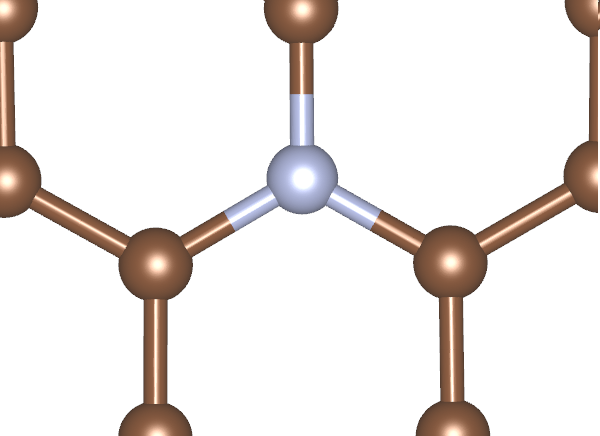} %
  } 
    \subfloat[\label{fig_1_d}]{%
  \includegraphics[width=0.3\columnwidth]{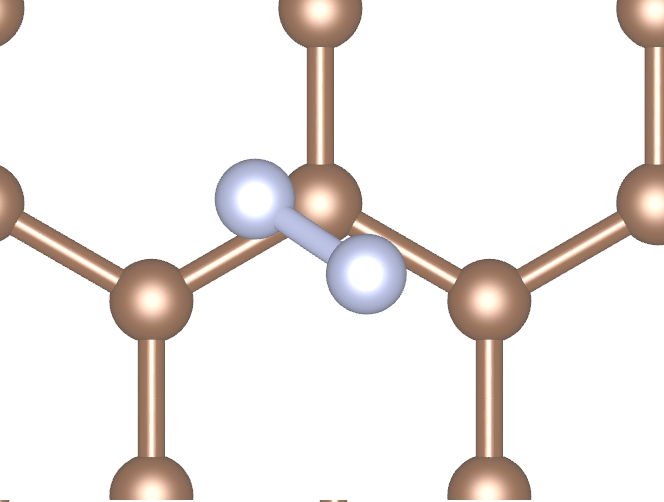} %
  } 
      \subfloat[\label{fig_1_e}]{%
  \includegraphics[width=0.45\columnwidth]{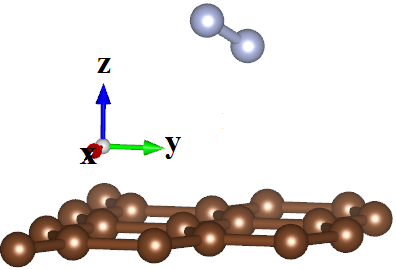} %
  } 
  \caption{ Atomic configurations for various impurity types. (a) Top and (b) side view of atomic adsorption, (c) top view of atomic substitution, and (d) top and (e) side view of molecular adsorption of nitrogen.}
  \label{fig_1}
\end{figure}

Experiments have shown that there are predominantly three types of N-graphene bonding, namely substitutional (graphitic), atomic adsorption (pyridinic and/or pyrrolic) and molecular adsorption \cite{exp_N_gr_lu2013nitrogen}. Fig. \ref{fig_1} shows the structures we explore in this work. Figs. \ref{fig_1_a} and \ref{fig_1_b} show the top and side views respectively of an adsorbed nitrogen atom in structures we denote by $XN50C$, where $X$ is the number of adsorbed nitrogen atoms per $(5\times5)$ supercell. In these structures, there is essentially no distortion of the lattice structure in the $x-y$ plane, but there is a significant movement of the carbon atoms towards the adsorbate in the direction perpendicular to the graphene sheet. Fig. \ref{fig_1_c} shows the top view of the atomic structure for the substitutional configuration. We denote such structures by $XN(50-X)C$. These structures exhibit essentially no displacement in the perpendicular direction relative to the structures with atomic adsorption. This is consistent with the previous DFT calculations of Mombrú et al. \cite{mombru2019electronic}. Figs. \ref{fig_1_d} and \ref{fig_1_e} show the top and side views respectively of the atomic structure for molecular nitrogen adsorbed on top of the graphene surface. This configuration is denoted by $1N_250C$.  For structures with more than one nitrogen atom per supercell, the positions of nitrogen atoms are chosen such that the adsorbates or substitutional atoms are located as far apart as possible while maximizing the symmetry of the structure. For example, for the $2N50C$ structure, one nitrogen atom is located above a $C$-$C$ bond that is oriented along the $x$ axis, while the other is above a bond that is oriented at $30^{\circ}$ to the $x$-axis. In what follows, We use $50C$ to denote results for pristine graphene obtained using SCC-DFTB for a relaxed $5\times 5$ unit cell and $2C$ to denote results for pristine graphene obtained using NNTB with a phenomenological Fermi velocity (and a two-carbon unit cell).\\
 
In Table \ref{table_1}, we present a number of the key structural parameters obtained from our SCC-DFTB calculations. First, we present the equilibrium distance of the adsorbed species to the graphene plane ($dist_N$) and to the first ($dist_{C1}$) and second ($dist_{C2}$) neighbouring carbon atoms. While substitutional and molecular adsorption of nitrogen causes negligible distortion, atomic adsorption of nitrogen causes a large disturbance in the lattice structure, such that after full relaxation, the nitrogen atom, and the first and second nearest neighbour carbon atoms to the adsorbed nitrogen are located 1.72 \text{\normalfont\AA} , 0.535 \text{\normalfont\AA}, and 0.2 \text{\normalfont\AA} above the plane of graphene, respectively. These distances are slightly reduced for higher surface coverage, as summarized in Table \ref{table_1}. Our results are consistent with the previous experimental results that show that substitutional nitrogen doping of graphene preserves the structure of the lattice, while atomic adsorption of nitrogen leads to significant lattice distortion \cite{exp_N_gr_lu2013nitrogen, jeong2011nitrogen_N_gr_fab, zhao2011visualizing_fab}. \\  
  


As shown in Table \ref{table_1}, we find that the nitrogen dopants are charged, such that an adsorbed nitrogen atom always receives some electronic charge from the surface,  while a substitutional nitrogen atom donates some electronic charge to the neighboring carbon atoms. Effective Mulliken charge differences ($\sigma_N$) are summarized in Table \ref{table_1}, where the charge difference is defined as the difference between the charge on a nitrogen atom in an isolated $N_2$ molecule and the charge on the nitrogen atoms bonded to the graphene. A negative sign corresponds to gain of electronic charge for the impurity atom. We also find that the charge difference per dopant decreases as the dopant density increases. These differences indicate the dependence on the limited size of the supercell.
Recent experiments indicate that with increasing doping concentration, N-doped graphene shifts from p-type to n-type, and an increasing electron-hole transport asymmetry arises \cite{exp_N_gr_lu2013nitrogen}. It is also known that graphene obtained by chemical vapour deposition is intrinsically p-doped while epitaxial graphene is intrinsically n-doped \cite{hafez2017Dopingeffects}. Our results show that substitutional N-doped graphene is n-type. This is consistent with  previous  studies \cite{schiros2012connecting,zhao2011visualizing_fab}. Moreover, previous DFT studies show that when nitrogen bonds with two carbon and interacts with a neighboring carbon vacancy (pyridinic or pyrollic) then the material is p type \cite{schiros2012connecting, usachov2011nitrogen_fab, zheng2010scanning_fab}. In our study, we have just simulated atomic N adsorbed to the surface and hybridizing with two carbons but no vacancy. However, the result is still p type.\\
\\ 

 \onecolumngrid  
 
\normalsize 
\begin{table}[h!]
\begin{center}
\begin{tabular}{| c | c c  | c c c| c |}
  \hline
  & $1N50C$ & $2N50C$ &  $1N49C$ & $2N48C$ & $3N47C$  & $1N_250C$,top  \\
  \hline
  $dist_{N}  $ (\text{\normalfont\AA}) & 1.726 & 1.692 &  0  & 0  & 0  &2.97 \\
  $dist_{C1} $ (\text{\normalfont\AA}) & 0.535 & 0.510 &  1.427  & 1.427  & 1.427  & 3.04  \\
  $dist_{C2} $ (\text{\normalfont\AA}) & 0.206 & 0.216 &  2.475  & 2.475  & 2.475 & 3.18 \\
  \hline
  $E_g$ (meV)& 0.4 & 84.1 &  95.7  & 174.8  & 237.3  & 0.7  \\ 
  $\sigma_N \times |e|$ & -0.38 & -0.34$\times 2$ & +0.125   & +0.1170$\times 2$  & +0.11638$\times 3$ & -0.0007  \\ 
    \hline
  $n_e/n_{2C} , \mu_c=100$ meV & 1.21  & 2.65  &  1.89   & 2.69   & 3.36   &  1.01 \\
  $n_e/n_{2C} , \mu_c=175$ meV & 1.22  & 2.26  &  1.61   & 2.15   & 2.59   & 1.01 \\  
      \hline 
\end{tabular}
 \caption{Atomic positions, band gaps, Mulliken charge differences and carrier densities for the six different dopant structures. The first three rows of the table give the equilibrium distances to the plane of graphene for the adsorbed species and out of plane displacement for the first and second neighbouring carbon atoms. The fourth and fifth rows give the band gap and induced effective Mulliken charge difference for the impurity atom, respectively. Note that a negative sign for effective Mulliken charge corresponds to gain of electronic charge on the nitrogen atoms. The final two rows give the ratio of the electron density at room temperature to that of the pristine graphene for two different chemical potentials (relative to the conduction band minimum). Each single defect per unit cell corresponds to a defect density of $7.56\times10^{13}\: \rm cm^{-2}$. }
 \label{table_1}
\end{center} 
\end{table}  

\twocolumngrid 

Molecular charge transfer doping with different molecular species (e.g. $NH_3$ as electron donor, $NO_2$ as electron acceptor) is an effective approach to realize an n or p type material  \cite{schedin2007detection_fab, zhang2011opening_fab}. The experiments of Docherty \textit{et al.} showed that molecular nitrogen can physisorb on graphene. As expected from the large equilibrium distance to the surface, we find (see Table \ref{table_1}) that adsorption of a $N_2$ molecule results in the smallest induced gross charge (Mulliken charge). This value is larger for the case when the $N_2$ is at the middle of the hexagon than when it is located on top of a surface carbon atom. As we shall see, this small charge transfer is not enough to significantly alter the band structure, and perhaps a much higher concentration of molecular doping is required to replicate the experimental findings of Docherty \textit{et al.} that the differential THz response of graphene changes significantly after exposure to nitrogen gas due to molecular adsorption of nitrogen. \\

Despite the differences in the intrinsic chemical potential of these structures, in what follows, we assume that gating has been used to set the chemical potential without perturbing the electronic band structure.  In Table \ref{table_1}, we show the calculated carrier densities at room temperature (found using the calculated energy bands) for the two different chemical potentials that we will consider in our dynamics calculations.  \\

\begin{figure}
 \centering
\subfloat[\label{fig_2_1}]{%
  \includegraphics[width=1\columnwidth]{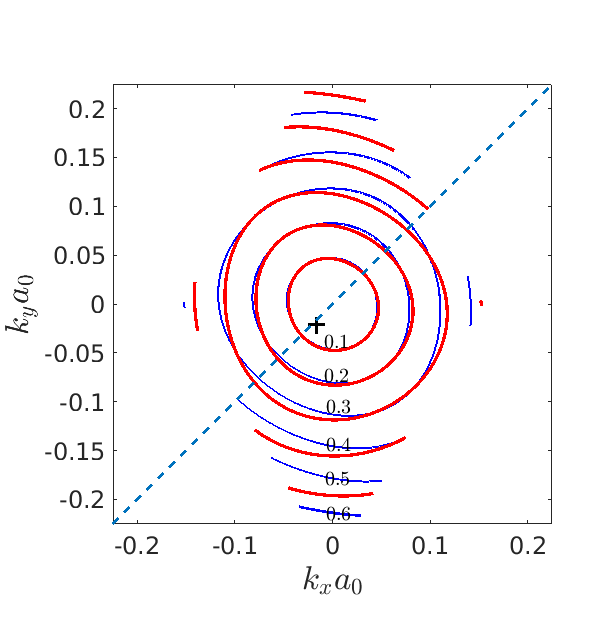}
} 
\hfill
 \subfloat[\label{fig_2_2}]{%
 \includegraphics[width=1\columnwidth]{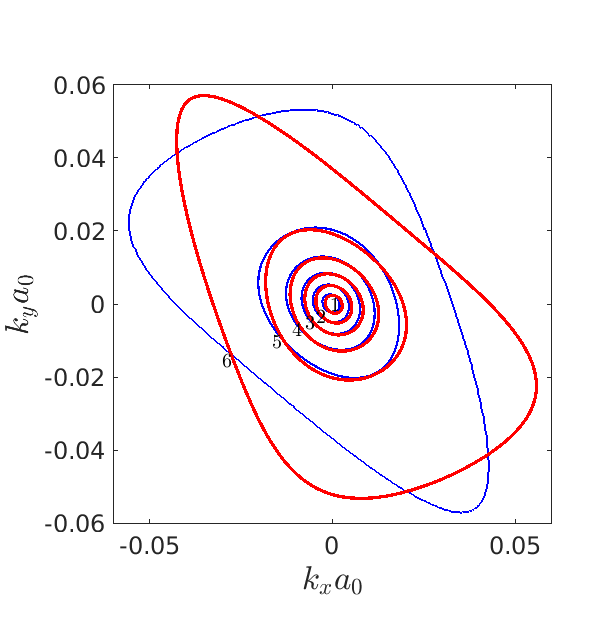}
} 
\caption{(a) Contour plot in $k$-space of the conduction band in the vicinity of the $K$ (blue) and $K^{'}$ (red) points for the $2N50C$ structure, where the contour values are in units of eV. Both contour plots are shifted such that the origin is at the position of the band minimum; the location of the $K$-point is indicated by a $+$. The dashed line shows the path in k-space ($k_x=k_y$) used for the plots presented in Fig. \ref{fig_3}. 
(b) Contour plot in $k$-space of the magnitude of the electron speed for the same structure, where the contour values are in units of $10^5$ m/s. Note that the area of $k$-space examined in (b) is much smaller than that of (a). 
} 
  \label{fig_2} 
\end{figure}   
 
\subsection{Energy bands, connection elements and Berry Curvatures} \label{res_energy_bands}

In this section we calculate the electronic band structure of graphene for different nitrogen dopant configurations and densities. The band gaps that we have calculated are given in Table \ref{table_1}. Note that significant band gaps open in all defected  structures except the one with molecular adsorption of nitrogen. Although the transferred charge and lattice distortion are both larger in structures with adsorbed rather than substitutional nitrogen (for the same doping density), substitutional nitrogen induces a larger band gap opening. For molecular adsorption, the N atoms are further from the surface and the induced band gap is very small and is of the same order as that found for the $50C$ pristine graphene structure. Some previous DFTB and DFT work on the band structure of graphene with defects reported a band gap opening at the nominal $K$-point of pristine graphene \cite{TB_dopedGraphene_leconte2010damaging}. DFT calculations of Nath et al. yield a band gap opening of approximately 300 meV for a nitrogen doping concentration of 5\% \cite{nath2014ab}, which is comparable to the band gap  of $237$ meV that we have calculated for 6\% substitutional coverage ($3N47C$ structure).

 Although we do find that impurities introduce a band gap, our detailed study of the whole Brillouin zone also shows that all defected structures show a shift of the conduction band minimum and valence band maximum away from the $K$ point of pristine graphene, except in the case of molecular adsorption, which yields a band structure similar to pristine graphene in all respects. As an example, in Fig. \ref{fig_2_1} we present a contour plot of the conduction band near the $K$ and $K{'}$ points for $2N50C$. For this structure, the band minimum is shifted approximately along the $k_x =k_y$ direction (dashed line in \ref{fig_2}). In addition, we see that unlike \textit{pristine}-graphene, the conduction band of defected structures does not exhibit rotational symmetry near the band extrema. This phenomenon is more evident in Fig. \ref{fig_2_2}, where we present the corresponding contour plot of the electron speed ($v= |\boldsymbol{\nabla_{k}} E_c(\mathbf{k})/\hbar|$) of the $2N50C$ structure near the $K$ and $K{'}$ points. Note also that as a result, the bands close to the $K$ and $K{'}$ points differ significantly. We must therefore account for the carrier dynamics at both $K$ and $K^{'}$ points simultaneously when solving the carrier dynamics equations. Among the studied structures, atomic adsorption of nitrogen result in the highest shift of the band extrema. We note that in all our calculations, we enforce $k$-space inversion symmetry of the bands that follows from time-reversal symmetry. This symmetry can be seen in Fig. \ref{fig_2} by comparing the results at $K$ and $K{'}$ (noting that $\mathbf{K}{'}=-\mathbf{K}$) \\

\begin{figure}
 \centering
 
 \subfloat[\label{fig_3_a}]{%
  \includegraphics[width=1\columnwidth]{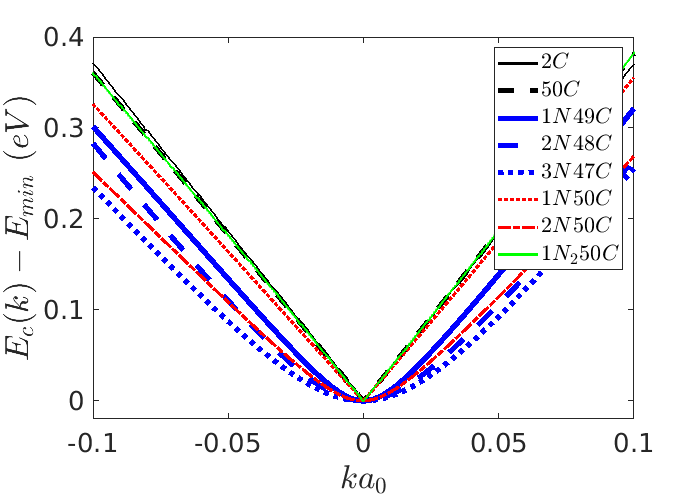}
} 
\hfill
 \subfloat[\label{fig_3_b}]{%
 \includegraphics[width=1\columnwidth]{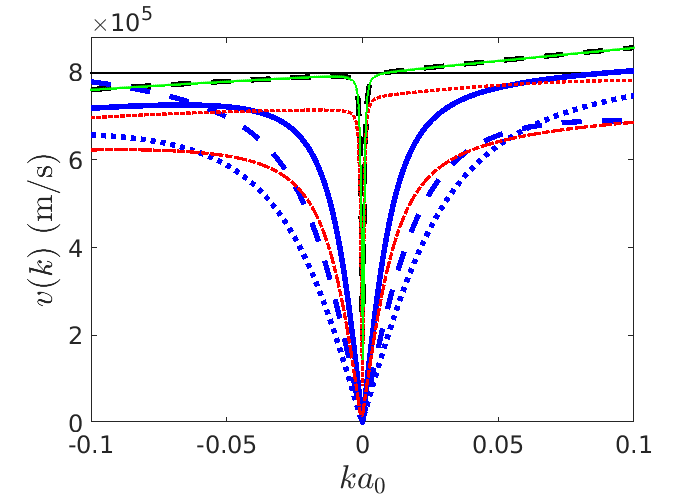}
} 
  \caption{ Cross sections of (a) the calculated conduction band structures and (b) carrier velocities for the  nitrogen doped structures summarized in Table \ref{table_1}.  The origin for each structure is set at the band minimum, which is shifted relative to the $K$ point.  Note that the dashed line in Fig. \ref{fig_2} shows the direction over which the 2D-cut is performed. For the analytic result (denoted by 2C), the Fermi velocity is set to $8.0\times 10^{5}$ m/s. }
  \label{fig_3}
\end{figure}

In Fig. \ref{fig_3_a} we plot a cross section of the calculated conduction bands and in Fig. \ref{fig_3_b} we plot the corresponding electron speed for eight different structures. The energy is relative to the band minimum for each structure and, as in Fig. \ref{fig_2}, the origin is at the band minimum near the $K$-point. The dashed line in Fig. \ref{fig_2} shows the direction in k-space used for the cross section. \\

We first consider pristine graphene. We have plotted the results from our SCC-DFTB calculation using a $(5\times5)$ cell ($50C$)as well as the analytic result for a NNTB calculation ($2C$) \cite{Marc_al2015nonperturbative, Marc_Ibr_2015optimizing} in which we have used a Fermi velocity of $v_f=8.0\times10^5 m/s$ to best match our computational bands. The slight relaxation of the $(5\times5)$ supercell has resulted in a deviation of the SCC-DFTB band structure from that of a perfect honeycomb structure. Consequently, our relaxed  graphene has bands that are not rotationally symmetric and has a band gap of 1.34 meV. Note, however that except very close to the band minimum, the band structure still has the essentially linear dispersion of a Dirac material. \\

We now consider the bands for the N-doped structures. We first note that the molecular adsorption structure ($1N_250C$) has a conduction band that is almost indistinguishable from that of pristine graphene. Fig. \ref{fig_3} shows that the structure with one adsorbed nitrogen ($1N50C$) exhibits characteristics of a Dirac-type material except very close to the Dirac point.  The substitutional structure ($xN(50-x)C$), however, has an almost parabolic band, even relatively far from the Dirac point. Note also that as the defect density increases for a given defect type, the bands become more parabolic. As we shall see, the degree of parabolicity plays a key role in the linear and nonlinear response of doped graphene.  \\

The relative parabolicity of the structures is more evident in Fig. \ref{fig_3_b}, where we plot the carrier speed, $v(k)$ as a function of $k$. We note first that the speed is noticeably lower for all of the doped structures than for pristine graphene. The velocity profile of the Dirac type structures is almost flat apart from a narrow dip at the symmetry point, while the velocity for structures with more parabolic bands exhibits a much wider dip and only flattens out when $ka_0\gtrsim 0.1$. For the same defect density, graphene with adsorption ($xN50C$) and substitution ($xN(50-x)C$) of atomic nitrogen exhibits the highest and the lowest carrier velocities for $k$ values close to the Dirac point, while the carrier velocity for the structure with molecular adsorption shows a velocity curve almost indistinguishable from that of $50C$ pristine graphene. Fig. \ref{fig_3_b} also shows that the relative ordering of the carrier speeds in the different structures is not the same for k-points further away from the Dirac-point as it is for points close to the Dirac point. As a result, changing the chemical potential can change which structures have the highest mobility. \\

\begin{figure}[h!]
 \centering
 \subfloat[\label{fig_4_a}]{%
  \includegraphics[width=.95\columnwidth]{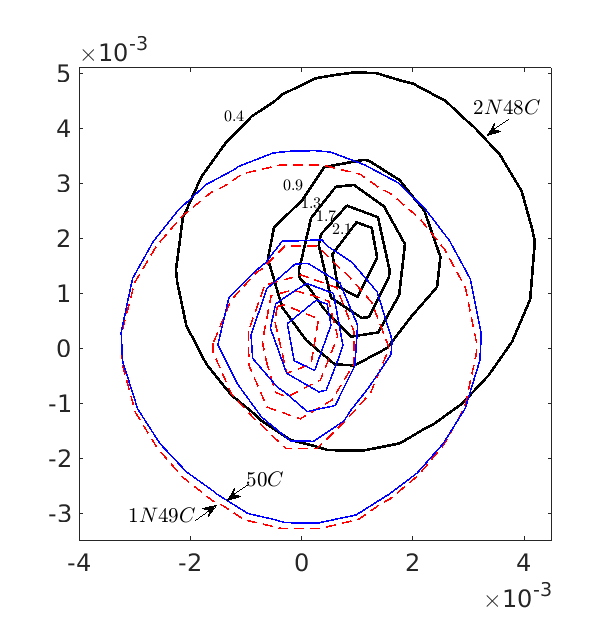}
} 
\hfill
 \subfloat[\label{fig_4_b}]{%
 \includegraphics[width=.95\columnwidth]{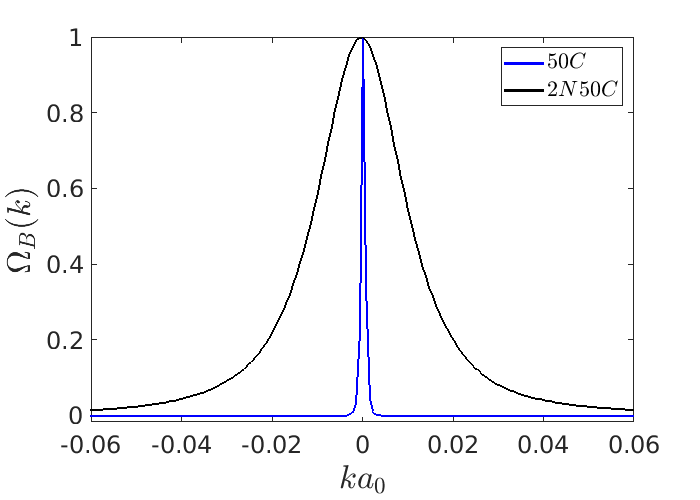}
} 
  \caption{ (a) Contour plot in $k$-space of the magnitude of the connection element of the conduction band $|\xi_{cc}(\mathbf{k})|$, for the $50C$, $1N49C$ and $2N48C$ structures. The contours are shifted in $k$-space such that the plot origin is at the position of the band minimum for each structure. The numbers on the contour lines for the $2N48C$ structure are in units of $10^{-8}$m. The Contour line values for the $1N49C$ structure are the same of that of the $2N48C$ structure, and are two times larger than contour line values of the $50C$ structure. (b) Cross section of the $z$-component of the calculated Berry curvature about the $K'$-point normalized to the peak value for $50C$ and $2N50C$ structures. See Fig. \ref{fig_2} for the direction over which the 2D-cut is performed.  }
  \label{fig_4}  
\end{figure} 
 
Using the Bloch states found from the SCC-DFTB calculation, we have calculated the intraband connection elements for the conduction band. Previous work \cite{Marc_al2015nonperturbative} has shown that in a NNTB calculation, the conduction-band connection elements of pristine graphene near the Dirac $K$-point are given by ($\boldsymbol{\xi}_{cc}(\textbf{K}+\textbf{k})= \hat{\boldsymbol{\theta}} / (2k)$, where $\hat{\boldsymbol{\theta}}$ is the angular unit vector in cylindrical coordinates with the origin at the Dirac point). Although the intraband connection elements are not gauge invariant (as they depend on the $k$-dependent phases chosen for the Bloch functions), the way in which we have chosen the phases when calculating the connection elements is the same for all the structures presented in Fig \ref{fig_4_a}.\\

We find that the connection elements of the defected structures are dependent on the defect type and density. The absolute value of the calculated connection elements of the conduction band are plotted in Fig. \ref{fig_4_a} for the $50C$, $1N49C$ and $2N48C$ structures. The contours have been shifted in $k$-space such that the origin is the location of the conduction band minimum. The connection elements of all the studied defected structures show a shift of the peak value away from the position of the band minimum, with the peak values and widths of the same order of magnitude as that of \textit{pristine} ($50C$) graphene.\\

Although the connection elements depend on the choice of the phases of the Bloch modes, the Berry curvature does not. In addition, it is the Berry curvature that appears in the expression for the conductivity (see Eq. (\ref{eq_15})), and so the Berry curvature is the more important quantity. It is easily shown that the Berry curvature for pristine ($2C$) graphene is zero everywhere, except at the Dirac points, where it is not defined. Due to symmetry breaking, we find that all of our structures have a non-zero Berry curvature near the Dirac points. The cross section of the $z$-component of the Berry curvature of the $50C$ and $2N50C$ structures about the $K{'}$ point are presented in Fig \ref{fig_4_b}. For the pristine $50C$ structure, the Berry curvature is essentially zero, except very close to the Dirac point.  It is non-zero due to the small lattice distortion that breaks inversion symmetry. We find that the Berry curvatures for the structures with substitutional nitrogen (not shown) are similar to that of the $50C$ structure. In contrast, the Berry curvature of the $2N50C$ structure is non-zero over a relatively large region of $k$-space (up to values of $k$ corresponding to band energies of about 100 meV).  \\

From Eq. (\ref{eq_15}), we see that the valley-Hall current around a given Dirac point is proportional to the integral in $k$-space of the Berry curvature multiplied by the occupation probability. We find that the magnitude of the integral in $k$-space of the Berry curvature over the regions close to the $K$ and $K{'}$ points for all structures is \rm $\pi$ (to within numerical precision).  However, because the Berry curvatures near the $K{'}$-point are the negative of those near the $K$-point, the net anomalous current density is always zero, as expected due to time-inversion symmetry. We note, however, that for a given defect type, the sign of the Berry curvature about each Dirac point is independent of the spatial location of the defect. Moreover, for structures with adsorbed nitrogen, inversion symmetry is broken, even when one averages over all possible defect locations. Thus, one expects a valley-Hall current in these defected structures that is absent in pristine graphene. If the carrier density is large enough such that all the states for which the Berry curvature is non-negligible are occupied, then the valley-Hall current due the to conduction band carriers in a given valley will have a magnitude of $e^2/(2\pi\hbar)$. Of course, a net anomalous current due to the Berry curvature will only arise if the carrier density around the $K$ point is different than that around the $K{'}$ point, which would be difficult to achieve in these structures, although schemes to accomplish this have been devised for other graphene systems \cite{Marc_friedlan2021valley,  Berry_shimazaki2015generation, Berry_farajollahpour2017role}. \\ 

\subsection{Linear and nonlinear response} \label{carrier dynamics}

\begin{figure}
 \centering
 
    \subfloat[\label{fig_5_a}]{%
 \includegraphics[width=1\columnwidth]{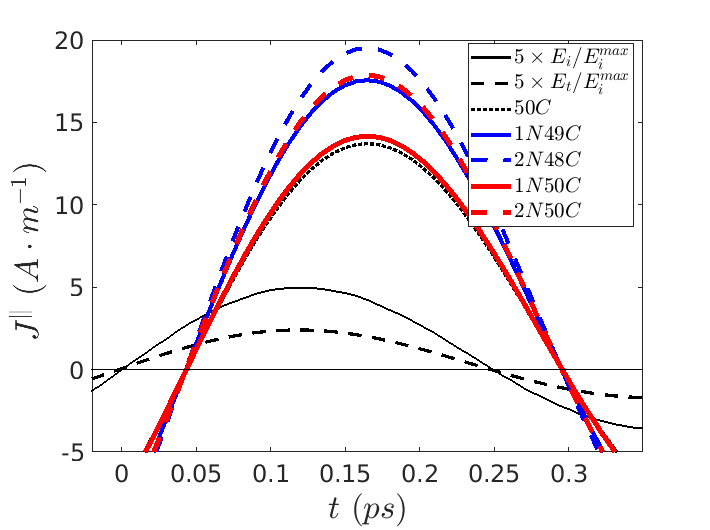}
} \hfill
    \subfloat[\label{fig_5_b}]{%
 \includegraphics[width=1\columnwidth]{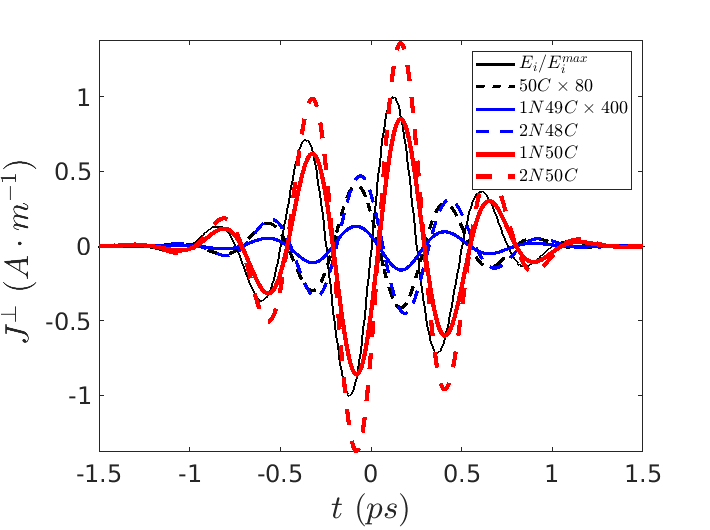} 
}

  \caption{ Current densities in the directions (a) parallel and (b) perpendicular to the incident field. $J^{\perp}$,$J^{\parallel}$, $E_i$, and $E_t$ indicate the current density in the perpendicular and parallel directions, and the normalized incident and transmitted field for the $50C$ structure, respectively. The incident and transmitted fields are plotted to highlight the phase shift of the current densities. Note that the timescales are very different in the two plots. 
  }
  \label{fig_5}
\end{figure}

In this section, we use the calculated band structures and connection elements in our density matrix formalism to compute the current density and transmitted electric field for an incident THz pulse. From these, we also calculate the linear and nonlinear mobility and the generated third harmonic signals. We do not present the results for the structure with adsorbed molecular nitrogen, as they are almost indistinguishable from those of pristine graphene. In all that follows, we assume that the chemical potential can be altered through gating without changing the equilibrium topology of the structure and the corresponding band structure. We run simulations at a temperature of $300$ K, for chemical potentials (relative to the conduction band minima) of $\mu_F$= 100 and 175 meV, corresponding, respectively, to carrier densities of $1.39 \times 10^{12}\: \rm cm^{-2}$ and $3.76 \times 10^{12}\: \rm cm^{-2}$  respectively in pristine graphene. The calculated carrier densities of the doped structures relative to the analytical carrier density of graphene at these chemical potentials are summarized in Table \ref{table_1}. The input THz field is a linearly-polarized, sinusoidal Gaussian pulse, with central frequency of $f_\circ =\rm{2}$ THz, and pulse width of 1 ps (FWHM) (see  Fig. \ref{fig_5_b}). Except where explicitly stated, the field is polarized in the $x$-direction. The phenomenological scattering time is taken to be $\tau_c=\rm{50}$ fs, independent of the impurity density or type.  This scattering time is used to account for various scattering mechanisms, such as neutral impurities, acoustic and optical phonons, and charged impurities in the substrate, which we assume are dominant over N-induced scattering \cite{Marc_Ibr_2015optimizing, Marc_Luke_2019effect}. We take the substrate to be silicon carbide, which has a refractive index of $n = 3$ at terahertz frequencies \cite{naftaly2016_SiC_refInd}. \\

Current densities, $J^\parallel (t)$ and $J^\perp (t)$  in the directions parallel and perpendicular to the incident field are shown in Figs. \ref{fig_5_a} and \ref{fig_5_b}, respectively. The normalized incident field is also plotted in Figs. \ref{fig_5_a} and \ref{fig_5_b} to highlight the phase of the current density relative to the incident field.  We only show the time interval close to $t=0$ for the parallel current so as to more clearly display the differences in the transmitted field amplitudes for the different doping configurations. From Fig. \ref{fig_5_a}, we see that all parallel current densities have essentially the same phase shift relative to the incident field but that the amplitudes differ considerably.  \\

In a NNTB calculation of pristine graphene, due to symmetry there is no current generated in the direction perpendicular to the incident field. However, due to the band symmetry breaking induced by the defects, this is not the case for N-doped graphene. We see from Fig. \ref{fig_5_b} that both the amplitude and the phase of the perpendicular current density relative to the incident field depend strongly on the directions of the nitrogen-carbon bonds relative to the electric field polarization. In all of our calculations, we have assumed particular locations for the defects within the supercell. These can be seen for the case of single defects in Fig. \ref{fig_1}. In a real structure, (assuming that most defects are well spaced out), the defects would be located with equal probability in the equivalent locations in the undisturbed lattice. If one calculates the average of the current response for the different locations, then the contribution to the linear part of the perpendicular current from the first term in Eq. (\ref{eq_15}) can easily be seen to average to zero; this is because this contribution to the linear conductivity tensor is symmetric under exchange of indices $x$ and $y$.  

\begin{figure}
 \centering

      \subfloat[\label{fig_6_a}]{%
 \includegraphics[width=1\columnwidth]{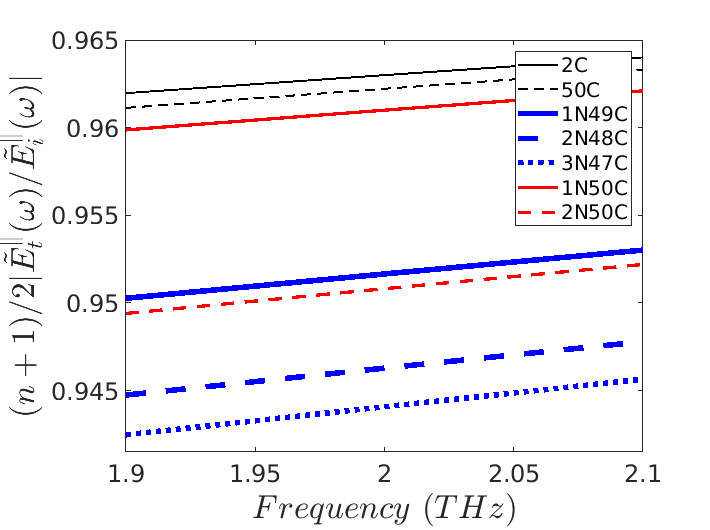}
}  \hfill
 
        \subfloat[\label{fig_6_b}]{%
 \includegraphics[width=1\columnwidth]{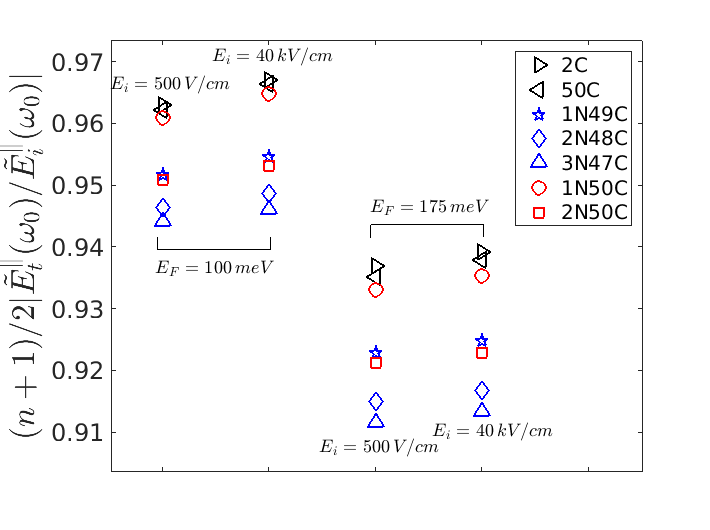}
} 

  \caption{ (a) The effect of defect type and density on field transmission for seven structures for  $f_0$=2THz, $\mu_c$=100meV and $E_i$=0.5 kV/cm. (b) The transmission values at the central frequency of $f_0$=2 THz for seven structures for the two different chemical potentials and incident field amplitudes indicated. The definition of transmission is chosen such that it will be 1 if there is no graphene. The tilde on the field indicates that it is the Fourier transform, while the superscript indicates that it is parallel to the incident field. }
  \label{fig_6}
\end{figure}

In Fig. \ref{fig_6_a} we plot the ratio of the transmitted field amplitude in the presence of graphene to the transmitted field amplitude in the absence of the graphene as a function of frequency for a chemical potential of 100meV, for a low (0.5 kV/cm) incident field amplitude. Here the tilde on the field indicates that it is the Fourier transform of the time-dependent field. As can be seen, the nitrogen doping  has a significant effect on the transmission and, as expected, the transmission decreases as the defect density increases for both adsorbed and substitutional doping.  \\

In Fig. \ref{fig_6_b} we present the transmission at the central frequency of 2 THz for the same structures for chemical potentials of 100 meV and 175 meV. When the chemical potential is increased, the carrier density and thus current density increases.  As a result, the transmission decreases significantly. The decrease in the transmission arises, of course, from an increase in the conductivity. Given that the average electron speed decreases when defects are added, it is at first sight surprising that the conductivity increases when nitrogen defects are introduced. However, as one can see from Fig. \ref{fig_3}, because the band shapes change significantly when defects are present, the carrier density for a given chemical potential is sensitive to both the defect type and density; this is seen clearly in Table \ref{table_1}. Thus, the decrease in the transmission when defects are introduced is due essentially due to an increase in carrier density.\\ 

 In Fig. \ref{fig_6_b}, we have also plotted the transmitted field amplitude at the central frequency for a higher incident field amplitude of 40 kV/cm. As can be seen, the transmission increases when the field is increased, indicating current saturation due to nonlinear effects. This increase in transmission with increasing field has been seen by experiments and theory by many authors \cite{hafez2014nonlinear, maeng2012gate, jnawali2013observation, winnerl2011carrier,Marc_Luke_2019effect}. \\ 
 
A better measure of the effect of defect type and density on the response of graphene to THz fields is the mobility, which we now examine. The mobility is defined generally by $\mu(\omega)= \sigma(\omega)/(e\,n_e)$, where $\sigma(\omega)$ is the surface conductivity and $n_e$ is the carrier density. Although the mobility in materials with parabolic bands is independent of the carrier density, this is not true for general band structures. In particular, the conductivity of pristine graphene in the NNTB approximation at temperature $T$ is given by \cite{drude_frenzel2014semiconducting, drude_shi2014controlling} 
\begin{equation}
\label{eq_17}
\sigma(\omega) =\frac{2e^2\ln{\left[2\cosh({\frac{\beta\mu_c}{2}}) \right]}}{\pi\beta \hbar^2 (1/\tau-i\omega)}   
\end{equation}
and the carrier density is given by
\begin{equation}
\label{eq_18}
n_{e}=-\frac{2}{\pi \left( \beta\hbar v_{F}\right)^2} Li_{2}\left(  -e^{\beta\mu_{c}}\right),
\end{equation}
where $\beta\equiv 1/k_{B}T$ and $Li_{2}\left(  z\right)  $, is the dilogarithm function.  From these equations, one finds that at zero temperature, the conductivity is proportional to the square root of the density and thus the mobility is proportional to $n_{e}^{-1/2}$. As we shall see, for the N-doped structures, the mobility also decreases with density, but not in exactly the same way. \\

\begin{figure}
 \centering 
 
 \subfloat{%
 \includegraphics[width=1\columnwidth]{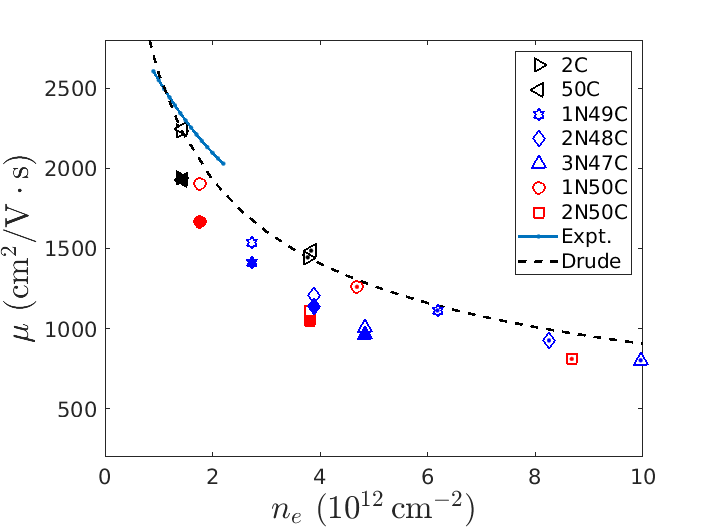}
} 
  \caption{Mobility versus carrier density for $E_i$= 0.5 kV/cm, $\mu_c$ = 100 meV (empty symbols) and 175 meV (symbols with dots), at the central frequency $f_0=2$ THz. Filled symbols represent the calculated mobility for $E_i$= 40 kV/cm, $\mu_c$ = 100 meV. The solid line represents the experimental results of graphene being exposed to various inert gases \cite{exp_knight2017situ}, the dashed lines show the calculated Drude-mobility for \textit{pristine} graphene from Eqs. (\ref{eq_17}) and (\ref{eq_18}). }
  \label{fig_7}
\end{figure}

In Fig. \ref{fig_7}, we plot the absolute value of the calculated mobility at the central frequency as a function of carrier density for different defect types and densities. For reference,  we have also plotted the curve for pristine graphene in the NNTB approximation. Note first that the analytic NNTB results agree with our 50C pristine graphene results to better than 0.5\%. Note also that for all densities, the mobility for the N-doped structures is lower than that of pristine graphene. In particular, we find that for the lower chemical potential of 100 meV, the mobility relative to Drude model for pristine graphene with the same carrier density is reduced by 8\% in structures with one defect per cell, by 15-22\% for structures with 2 defects per cell, and by 21\% for the structure with 3 defects per unit cell. For the higher chemical potential of 175 meV, the decreases in mobility are somewhat less: the mobility is reduced by 3\% for structures with one defect per cell, by 7-16\% for structures with 2 defects per cell, and by 11\% for the structure with 3 defects per unit cell. \\ 

The reduction in the defect-induced degradation in the mobility with increasing chemical potential arises largely from the dependence of the carrier velocity on energy. When the chemical potential reaches an energy at which the carrier velocities have reached their maximum, the effect of N-doping on the mobility is modest (see Fig. \ref{fig_3_b}). For example, at the lower chemical potential, the $2N50C$ structure has a velocity comparable to that of $2N48C$ structure, which results in its mobility being close to that of $2N48C$. However, as the chemical potential increases, the velocity and hence the mobility of the $2N50C$ becomes comparable to that of $3N47C$. \\  


Knight \textit{et al.} \cite{exp_knight2017situ} have performed in-situ terahertz optical Hall effect experiments to extract the mobility of N-doped graphene as a function of carrier density. The experiments were carried out for different exposure times to $\rm He$, air, and $\rm N_{2}$ for a central frequency of 0.428 THz
and yielded similar results for the mobilities for the different gasses. In Fig. \ref{fig_7} we have included the average of the experimental results (solid blue line). As can be seen, our results are generally in good agreement, falling only 13\% below the experimental results. A lower value for our results is expected because of the lower frequency at which the experiments are done. However, a direct comparison is not possible because the experimental defect density is unknown, and changes as the carrier density changes. \\

       
We now turn to examine the effects of N-doping on the dependence of the mobility on the field amplitude. In Fig. \ref{fig_7} we have included calculated mobilities for pristine graphene. At the lower field strength, our results match the Drude mobility calculated in the NNTB approximation. At the higher THz field amplitude of 40 kV/cm, the mobilities are reduced in all cases, although the reduction is smaller in the nitrogen-doped samples than in pristine graphene. Furthermore, the field-induced reduction in the mobility is smallest in structures whose bands are more parabolic. We obtain a reduction in the mobility of about 13\% in \textit{pristine} graphene, about 8-12\% in structures with one defect per cell, about 6\% for structures with 2 defects per cell, and 4.5\% for the structure with 3 defects. The reduction in mobility with increasing field amplitude arises from the nonlinear response of the material. In pristine graphene, the relatively strong nonlinear response is a result of the linear bands, which results in a large $\chi^{(3)}$ response \cite{wright2009strong_chi_3_graphene, ishikawa2010nonlinear_chi_3_graphene}. Thus, again the relatively smaller effect of the field amplitude on the mobility for some of the structures (particularly the $2N48C$, $3N48C$, and $2N50C$ structure) is due to the fact that their bands are more parabolic and thus their $\chi^{(3)}$ is expected to be smaller.\\


\begin{figure}
 \centering
   \subfloat[\label{fig_8_a}]{%
 \includegraphics[width=1\columnwidth]{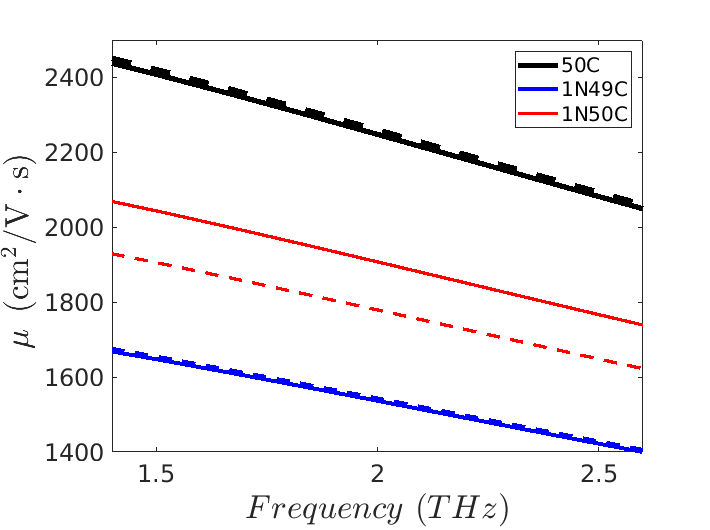}
} \hfill

\subfloat[\label{fig_8_b}]{%
 \includegraphics[width=1\columnwidth]{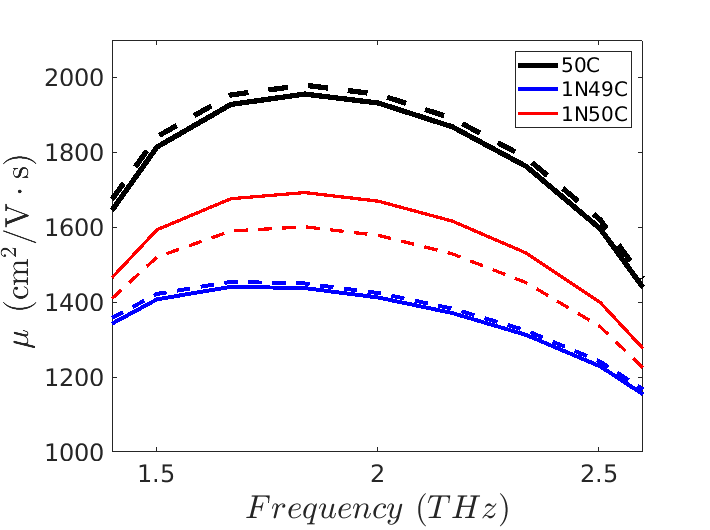}
} 
     \caption{Effect of the polarization of the incident field on the mobility spectrum for $\mu_c = 100$ meV for (a) $E_i = 0.5$ kV/cm  and (b) $E_i = 40$ kV/cm for three different structures. Dashed and solid lines correspond to the incident field polarized in the $y$ and $x$ directions, respectively. }
     \label{fig_8}
\end{figure}

Fig. \ref{fig_8} shows the effect of the polarization of the incident field on the mobility spectrum for a few different structures for low and high incident THz fields. In the previous section it was shown that defects remove the rotational symmetry near the Dirac points and this affects the carrier distribution, velocity and connection elements, 
and thereby the current density. The overall conductivity tensor is a function of the topology of the introduced defects. The difference in the calculated mobility of the same structure for fields that are polarized in the $x$ and $y$ directions is due entirely to the asymmetry in the band structure. As can be seen, the dependence on the field polarization is relatively small, with at most a 5\% change in the mobility. In a real structure, this anisotropy would average out and the final parallel mobility would be an average of the mobilities found for the two different polarizations. We have performed similar calculations for the higher incident field amplitude of 40 kV/cm (Fig. \ref{fig_8_b}) and find a similar anisotropy. \\

We now turn to another result of the nonlinear response of graphene: third harmonic generation (THG). Harmonic generation in graphene at terahertz frequencies, has been studied theoretically \cite{Marc_Ibr_2015optimizing, ishikawa2010nonlinear_chi_3_graphene, wright2009strong_chi_3_graphene, hong2013optical} and experimentally \cite{bowlan2014ultrafast, HG_kovalev2017probing}. Since the electron velocity of pristine graphene is independent of the crystal momentum near the Dirac points, a strong incident electric field shifts the majority of electrons to one side of the Dirac point in k-space, which causes current saturation and odd harmonic generation in the transmitted electric field \cite{Marc_Ibr_2015optimizing}. Because the carrier speed is no longer constant in N-doped structures, we expect the nonlinear response to be different in such structures. We observed this effect in the nonlinear transmission; we now examine its effect on harmonic generation.

\begin{figure} 
 \centering
\subfloat[\label{fig_9_a}]{%
 \includegraphics[width=1\columnwidth]{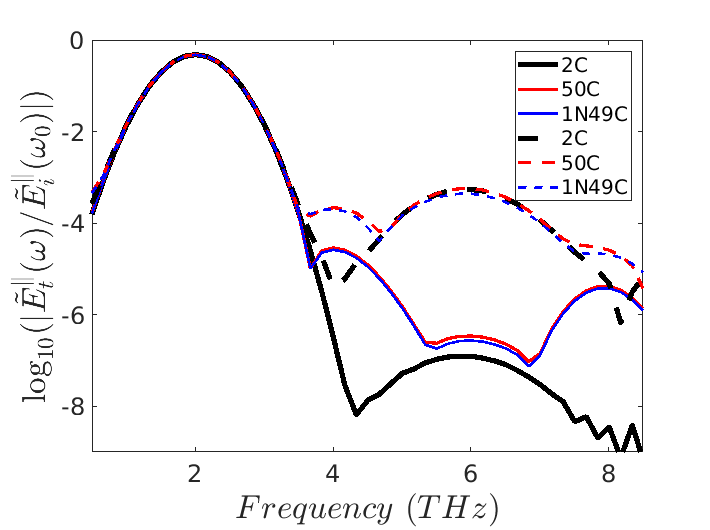}
} \hfill
\subfloat[\label{fig_9_b}]{%
 \includegraphics[width=1\columnwidth]{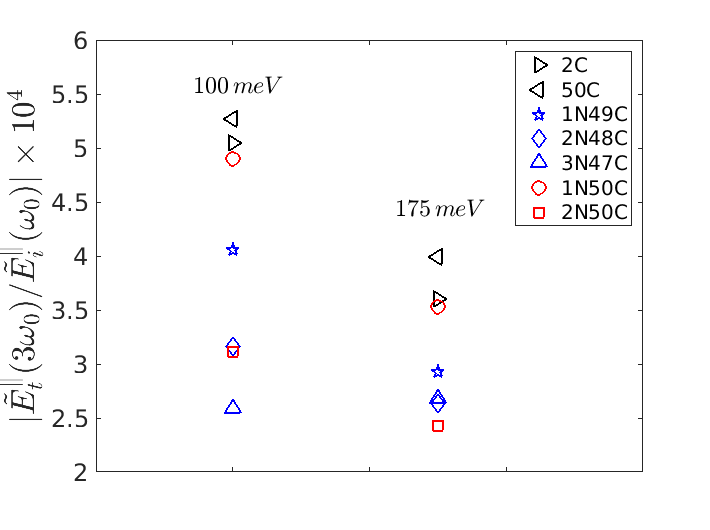}
}
  \caption{Effect of defect density, chemical potential and incident field on third harmonic generation. (a) The normalized transmitted field as a function of frequency for the six structures indicated for $\mu_c = 100$ meV, for incident field amplitudes of $E_i = 0.5$ kV/cm (solid) and $E_i = 40$ kV/cm (dashed). (b) Third harmonic field amplitude at 6 THz relative to the incident field at the fundamental for seven structures for $E_i = 40$ kV/cm, for $\mu_c= 100$ meV (left column), and $\mu_c=175$ meV (right column). The tilde denotes the Fourier transform of the field, while the superscript denotes the component parallel to the incident field. }
  \label{fig_9}
\end{figure}

The calculated spectrum of the transmitted electric field $\tilde{E}_t(\omega)$ normalized to the maximum amplitude of the incident field $\tilde{E_i}(\omega_0)$ is shown in Fig. \ref{fig_9} for three different structures and for two different incident THz field amplitudes. The peak at 2 THz is the fundamental, the peak seen at 4 THz for all but 2C graphene is the second harmonic, and the peak at 6 THz is the third harmonic. For the lower incident field amplitude of 0.5 kV/cm, the magnitude of generated third harmonic is approximately 6 orders of magnitude smaller than that of the fundamental, and so is negligible. On the other hand, for the higher incident field of 40 kV/cm, the third harmonic signal is down by less than three orders of magnitude and is thus experimentally observable \cite{hafez2020terahertz_THG, hafez2018extremely}. Interestingly, the spectrum of the defected structures (and even the 50C structure) shows a peak at the second harmonic that is above that of the third harmonic for the lower  incident field. Second harmonic generation arises from the asymmetry of the relaxed supercell and is not seen in the results for the 2C band structure. Note, however, that the second harmonic is lower than the third harmonic for the higher incident field. The randomized distribution of defects in a macroscopic sample would be expected to restore inversion symmetry on average, and thus a second harmonic signal would not be seen. These second harmonic peaks should therefore be viewed as artifacts of our particular choice for defect locations. \\

In Fig. \ref{fig_9_b} we plot the third harmonic amplitude for the different structures for the two different chemical potentials. As expected, we find that the third harmonic signal is reduced as the defect density is increased. Moreover, for the same defect density, the structures with a more parabolic band have lower THG. In other words, as expected, the bands that are most Dirac-like have higher THG. Interestingly, when the chemical potential is raised, the third harmonic signal decreases for all structures except for the 3N47C structure. In a previous work \cite{Marc_Ibr_2015optimizing}, we showed that for pristine graphene, there is an optimal chemical potential that maximizes the THG for a given scattering time. For a 1 THz pulse, we found that the optimal chemical potential was approximately 165 meV. Thus, it is not unexpected that the third harmonic signal will decrease when we increase the chemical potential from 100 meV to 175 meV for structures that have similar bands to pristine graphene, which is what we observe.  However, for the 3N47C structure, we see an increase in the THG signal when the doping is increased.  For this structure, as we have seen, the conduction band is essentially parabolic at low energies and only becomes essentially linear at higher energies. This transition occurs in the energy range between 120 meV and 200 meV.  Thus for the lower chemical potential of 100 meV, the nonlinearity is very low, as the electrons are largely in the parabolic band regime, while when the chemical potential is raised to 175 meV, many of the electrons are in the linear-dispersion regime, which yields a stronger nonlinearity.

\section{Conclusion} \label{conclusion}

In this work we have presented a detailed description of the band structure and linear and nonlinear THz response of nitrogen-doped graphene. Our calculations are in general agreement with the previous experimental and computational results that indicated that nitrogen doping causes band gap opening and semiconducting characteristics, and alters carrier density and carrier mobility. \\


Detailed study of the whole Brillouin zone showed that all defected structures exhibit a shift of Dirac points from their nominal points in \textit{pristine} graphene, and that there is an induced asymmetry between the $K$ and $K^{'}$ points.  Our results also show that in general, N-doping transforms \textit{pristine} graphene from a Dirac-type to a more parabolic-type material (near band minimum), and that the bands generally become more parabolic as the defect density increases. For the studied defect types and topologies, the calculated band structure and the associated linear and nonlinear terahertz response is sensitive both to defect density and defect type. We found that for the same defect density, the substitutional structure has the most parabolic band of all. In addition, as the defect density increases for a given defect type, the bands become more parabolic.\\




We used our model to calculate the mobility of the N-doped structures for frequencies close to 2 THz, with the largest changes occurring for structures that have bands that are more parabolic. We find qualitative agreement with recent experiments \cite{exp_knight2017situ}. We also found that the mobility
of all of the doped systems decreased with increasing incident THz field amplitude, but that the change was smallest when the conduction bands were more parabolic. This is due to the fact that the intrinsic nonlinearity in graphene arises from the band nonparabolicity. Consistent with this, we found that the generated third harmonic is reduced due to doping; for substitutional doping at the higher doping levels the reduction is almost a factor of two.  \\


Finally, we found that Berry curvatures of the defected structures are non-zero and highly dependent on the defect type and density. Although the macroscopic isotropy of the structures ensures that the conductivity tensor is diagonal, our study predicts a valley-Hall effect in the structures with adsorbed atomic nitrogen that is absent in pristine graphene.\\

In summary, we have found that the linear and nonlinear mobilities of graphene depend significantly on the type and density of nitrogen doping, with the key predictor of the mobility being the degree of parabolicity of the conduction band near the Dirac points. In future work, we plan to examine in detail the valley-Hall effect in doped graphene and to consider the linear and nonlinear carrier dynamics in systems with other dopants, specifically water and oxygen. \\

\textbf{Acknowledgement}\   

This work was supported by the Natural Sciences and Engineering Research Council of Canada (NSERC) and Compute Canada.


\appendix
\section{Bloch State Normalization} \label{Appendix A}
\renewcommand{\theequation}{A.\arabic{equation}}
\setcounter{equation}{0}
In this appendix, we derive the conditions on the Bloch state expansion parameters to ensure that the states are normalized. From Eq. (\ref{eq_6}) we know that normalization requires 
\begin{align}
\label{eq_A_1}
\begin{split}
\int_{\Omega_0} d^3\textbf{r}\, u^*_{n\textbf{k}}(\textbf{r})u_{m\textbf{k}}(\textbf{r}) & = \frac{\Omega}{(2\pi)^2} \delta_{nm},    \\ 
\end{split}
\end{align}
where $\Omega_0$ is the cell volume at ${\textbf{R}}=0$ and $\Omega$ is the 2D cell area. Now applying Eqs. (\ref{eq_1}) and (\ref{eq_2}) results in 
\begin{align} 
\begin{split}
& \int_{\Omega_0} d^3\textbf{r}\, u^*_{n\textbf{k}}(\textbf{r})u_{m\textbf{k}}(\textbf{r})  =\\
&\sum_{ \textbf{R}, \textbf{R}^{'}} \sum_{jj^{'}} C^{n*}_j (\textbf{k}) C^{m}_{j^{'}} (\textbf{k})
e^{i\textbf{k}\cdot(\textbf{R}^{'}-\textbf{R})} \\
& \times \int_{\Omega_0} d^3\textbf{r} \phi^*_j(\textbf{r} - \textbf{r}_j-\textbf{R}) \phi_{j^{'}}(\textbf{r} - \textbf{r}_{j^{'}}-\textbf{R}^{'}) \\
& = \sum_{ \textbf{R}, \textbf{R}^{''}} \sum_{jj^{'}} C^{n*}_j (\textbf{k}) C^{m}_{j^{'}} (\textbf{k})
e^{i\textbf{k}\cdot\textbf{R}^{''}} \\
& \times \int_{\Omega_{\textbf{R}}} d^3\textbf{r} \phi^*_j(\textbf{r} - \textbf{r}_j ) \phi_{j^{'}}(\textbf{r} - \textbf{r}_{j^{'}}-\textbf{R}^{''}) \\
& = \sum_{ \textbf{R}} \sum_{jj^{'}} C^{n*}_j (\textbf{k}) C^{m}_{j^{'}} (\textbf{k})
S(j,j^{'},\textbf{R}) e^{i\textbf{k}\cdot\textbf{R}},\\ 
\end{split}
\end{align}
where $\Omega_{\textbf{R}}$ is unit cell centred at $\textbf{r} = \textbf{R}$, and the overlap integral $S$ is defined as  
\begin{equation}
S(j,j^{'},\mathbf{R}) = \int_V d^3\textbf{r}  \phi^*_j(\textbf{r} - \textbf{r}_j ) \phi_{j^{'}}(\textbf{r} - \textbf{r}_{j^{'}}-\textbf{R}),
\end{equation}
where $V$ is the volume of all space. 
Therefore, Eq. (\ref{eq_A_1}) can be written as 
\begin{equation}
\sum_{jj^{'}} C^{n*}_j(\textbf{k}) C^{m}_{j^{'}}(\textbf{k}) S_{jj^{'}}(\textbf{k}) = \frac{\Omega}{(2\pi)^2}\delta_{nm}, 
\end{equation}
where 
\begin{equation}
\label{eq_A_5}
S_{jj^{'}}(\textbf{k})=\sum_{\textbf{R}} S(j,j^{'},\textbf{R}) e^{i\textbf{k}.\textbf{R}}.
\end{equation}


\section{Connection Elements } \label{Appendix B}
\renewcommand{\theequation}{B.\arabic{equation}}
\setcounter{equation}{0}
In this appendix we derive an analytic expression for connection elements when the overlaps are not zero. Calculating the approximate gradient of Eq. (\ref{eq_7}), by neglecting the gradient of the exponential term (which can be shown to be negligible \cite{Marc_al2015nonperturbative}), and using Eq. (\ref{eq_1}) in  Eq. (\ref{eq_11}), we obtain
\begin{align}
\begin{split}
& \boldsymbol{\xi}_{nm}(k) =\frac{i(2\pi)^2}{\Omega} \int_{\Omega_0} d^3r u^*_{n\textbf{k}}(\mathbf{r}) \boldsymbol{\nabla}_{\textbf{k}} u_{m\textbf{k}}(\mathbf{r})    \\
& \approx \frac{i(2\pi)^2}{\Omega} \sum_{ j,j^{'}} C^{n*}_{j}(\textbf{k})  \left[\boldsymbol{\nabla}_\mathbf{k}  C^m_{j^{'}} (\textbf{k})\right] \sum_{ \textbf{R}, \textbf{R}^{'}} e^{-i\textbf{k} \cdot (\textbf{R} - \textbf{R} ^{'})} \\
& \times \int_{\Omega_0} d^3\textbf{r} \phi^*_{j}(\textbf{r}-\textbf{r}_{j} -\textbf{R}) \phi_{j^{'}}(\textbf{r}-\textbf{r}_{j^{'}} -\textbf{R}^{'}) \\
& = \frac{i(2\pi)^2}{\Omega} \sum_{ j,j^{'}}C^{n*}_{j}(\textbf{k})  \left[\boldsymbol{\nabla}_\mathbf{k}  C^m_{j^{'}} (\textbf{k})\right] \\
& \times \sum_{ \textbf{R}, \textbf{R}^{''}} e^{i\textbf{k} \cdot   \textbf{R} ^{''}} \int_{\Omega_{\textbf{R}}} d^3\textbf{r} \phi^*_{j}(\textbf{r}-\textbf{r}_{j} ) \phi_{j^{'}}(\textbf{r}-\textbf{r}_{j^{'}} -\textbf{R}^{''})\\
& = \frac{i(2\pi)^2}{\Omega} \sum_{jj^{'}}  C_{j}^{n*}(\textbf{k})  \left[\boldsymbol{\nabla}_\mathbf{k}  C_{j^{'}}^{m}(\textbf{k})\right]
S_{jj^{'}}(\textbf{k}),  
\end{split}
\end{align}
which is the expression we have in Eq. (\ref{eq_12}).  \\

\bibliographystyle{apsrev4-2}
\bibliography{THz}

\end{document}